 \documentclass[11pt,5p,times,twocolumn]{elsarticle}
 
\usepackage{amssymb}
\usepackage{lipsum}
\usepackage{braket}
\usepackage{amsmath}
\usepackage{mathtools}
\usepackage{xcolor}
\usepackage{float}
\usepackage{comment}
\usepackage{amsthm}
\usepackage{url}
\usepackage{xurl} % Allows URLs to break at any character
\usepackage[colorlinks=true,allcolors=blue]{hyperref}
\usepackage{subcaption}
\usepackage{booktabs}
\usepackage[table]{xcolor}

\usepackage{lineno}

\biboptions{numbers,sort&compress}

\newcommand{\planckmass}{M^2_{pl}}

\newcommand\joao[1]{\textcolor{blue}{[{\bf Joao:} #1]}}
\newcommand\kaynan[1]{\textcolor{orange}{[{\bf Kaynan:} #1]}}

\journal{Physics Letters B}

\begin{document}

\begin{frontmatter}



\title{
    Beyond $w_0-w_a$ and Phantom Crossing: Testing Models of Coupled Dark Sector
}


\author[a]{Kaynan R. de O. Pompeu}
\ead{kr.pompeu@unesp.br}
\author[b]{João Rebouças}
\ead{joaoreboucas@arizona.edu}
\author[a]{Rogerio Rosenfeld}
\ead{rogerio.rosenfeld@unesp.br}
\affiliation[a]{
            organization={Instituto de Física Teórica da Universidade Estadual Paulista and ICTP South American Institute for Fundamental Research},
            addressline={R. Dr. Bento Teobaldo Ferraz, 271, Bloco II, Barra Funda}, 
            city={São Paulo},
            postcode={}, 
            state={SP},
            country={Brazil}}
\affiliation[b]{organization={Department of Astronomy/Steward Observatory, University of Arizona},
            addressline={933 North Cherry Avenue}, 
            city={Tucson},
            postcode={87521}, 
            state={AZ},
            country={USA}}

\begin{abstract}
The combination of cosmic microwave background (CMB) measurements with distance measurements from baryon acoustic oscillations (BAO) and type Ia supernovae (SNIa) suggests that dark energy is dynamical, with an equation of state crossing the phantom divide at low redshifts. This feature can not be described within canonically normalized, minimally coupled, self-interacting scalar field models. We investigate the possibility of achieving phantom crossing by introducing an interaction in the dark sector such that the dark matter particle mass is dependent on the dark energy field $\phi$ as $m \propto \phi^\alpha$. We consider a self-interacting potential of the inverse power-law form $V \propto \phi^{-\beta}$. We implement this model both at the background and perturbative levels in a Boltzmann solver and use a bayesian framework 
to constrain its parameters using using CMB, BAO and SNIa data. We find best-fits that have similar goodness-of-fits as the commonly used phenomenological $w_0-w_a$ parameterization, providing a more fundamental understanding of the dark sector.
\end{abstract}


\end{frontmatter}

%\tableofcontents

%\linenumbers

\section{Introduction}
\label{sec:introduction}

The fundamental nature of dark energy (DE) and cold dark matter (CDM) remains unknown, despite their existence being strongly supported by a wide range of cosmological observations. In the highly successful standard cosmological model $\Lambda$CDM, DE is treated as the manifestation of a simple cosmological constant $\Lambda$ \cite{Turner:2022gvw}, whose equation of state is $w=-1$ identically. It has become common to test $\Lambda$CDM against a model-agnostic extension where, instead of a cosmological constant, the dark energy is described by a fluid with an equation-of-state $w$ that evolves with time with the so-called $w_0-w_a$ (or Chevallier-Polarski-Linder, CPL) parameterization \cite{Chevallier:2000qy,Linder:2002et},
\begin{equation}
w(a) = w_0 + w_a (1-a),
\end{equation}
where $a(t)$ is the scale factor of the Universe. This extension belongs to the class of smooth dark energy models, where dark energy has negligible inhomogeneities, and the large-scale structure formation is only altered through the background expansion modification~\cite{Mortonson2009, Reboucas2024}.

There has been recent evidence that this simple $w_0-w_a$ extension is favored in comparison to $\Lambda$CDM to describe cosmological observations. The Dark Energy Spectroscopic Instrument (DESI) collaboration, when combining its measurements of the Baryon Acoustic Oscillation (BAO) with type Ia supernova (SN) distances and cosmic microwave background (CMB) anisotropies, claims that $\Lambda$CDM is rejected with statistical significance of approximately 3$\sigma$, depending on the SN dataset used~\cite{Adame_2025, Calderon_2024, lodha2025extendeddarkenergyanalysis, Gu_2025, Abdul_Karim_2025, union31}.
More recently, the Dark Energy Survey (DES) collaboration presented results on the $w_0-w_a$ extension using the final dataset of its six years of observations \cite{DESExtensions}, obtaining a 2.2$\sigma$\, deviation from a cosmological constant using a combination of galaxy correlation functions, supernovae luminosity distances, and BAO distances, the tightest constraints ever obtained from a single survey to date. These unexpected results might indicate the first hints of dynamical dark energy, and will be further elucidated by Stage-IV surveys such as Euclid~\cite{euclid_overview}, the Legacy Survey of Space and Time (LSST)~\cite{lsst-srd}, and the Roman Space Telescope~\cite{roman_wfi, roman_multiprobe}, as well as precise CMB measurements from the Simons Observatory~\cite{simons}. 

One of the intriguing aspects of this still unsettled evidence for deviations of $\Lambda$CDM is that, using this $w_0-w_a$ extension, the best-fit parameters imply that the dark energy equation-of-state becomes phantom (\textit{i.e.} $w < -1$)  in the past, with a crossing to this phantom regime at $z\approx0.5$~\cite{reboucas2025investigatinglatetimedarkenergy}. This is puzzling because fundamental models describing dark energy usually rely on a scalar field rolling down a given potential. Such models are commonly referred to as quintessence models (for reviews, see {\it e.g.} \cite{Amendola2010} and \cite{Tsujikawa_2013}), and with a non-interacting, minimally coupled scalar field with a canonical kinetic term, a phantom behavior cannot be achieved \cite{nojiri2026phantomdividecrossinggeneral}.

This puzzle has led to a large body of recent papers proposing more complex dark energy scalar field models that can address the phantom crossing phenomenon. These include, among many other theoretical possibilities: non-minimally coupled models \cite{Wolf_2025, Ye_2025, Pan_2026, Wang_2026, Wang_2026_2, nojiri2026apparentphantomcrossinggaussbonnet, nojiri2025phantomcrossingoscillatingdark}, models with non-canonical kinetic terms (also known as k-essence) \cite{Goldstein:2025epp}, particle-physics models that are consistent with naturalness \cite{Delaunay:2026jto,Delaunay:2026fse}, and coupled dark sector (CDS) models which assume interactions between dark energy and dark matter \cite{Miranda2018, Khoury:2025txd,Chakraborty_2025,li2026strongevidencedarksector,Chakraborty_2026,Guedezounme_2026,lapenna2026mimickingphantomdarkenergy,gomezvalent2026constraintscoupleddarkenergy,chen2025quintessentialdarkenergycrossing,liu2026phantommirageaxiondark,antusch2026coupleddarkenergydark, Petri_2026},
to name a few.

The coupled scenario is particularly attractive because it can be realized using simple fundamental models with canonical fields, evading theoretical difficulties that arise in some of the alternative models.
In particular, it was shown long ago by one of the authors that this class of models can lead to an apparent phantom crossing~\cite{Rosenfeld_2007} (see also \cite{Das_2006, PhysRevD.80.123001}), a phenomenon that was recently dubbed ``phantom mirage" \cite{liu2026phantommirageaxiondark}.
These models can be fully described by a scalar field potential $V(\phi)$ and by the scalar field-dependent dark matter mass $m(\phi)$ arising from the dark energy-dark matter interaction.  In this work, a phenomenologically motivated self-interacting potential for the DE component proposed by Peebles and Ratra \cite{Peebles:1987ek, Ratra:1987rm} is adopted, $V(\phi) \propto \phi^{-\beta}$, and a power-law form is assumed for $m(\phi)\propto \phi^{\alpha}$ (for a different choice see, \textit{e.g.} \cite{Turyshev:2026yye}). This model reduces to $\Lambda$CDM for $\alpha=\beta=0$.

We constrain the parameters of this CDS model using measurements of CMB anisotropies, SN luminosity distances, and distances inferred using the BAO feature in galaxy correlation functions in a Bayesian framework. Furthermore, we assess the goodness-of-fit of this model by finding the maximum a posteriori (MAP), and comparing the results against the common $w_0-w_a$ parameterization. We find that this CDS model can explain current data at the same level of the $w_0-w_a$ parameterization, providing a more fundamental understanding of dark energy.

%We split it into two different types and, by selectively increasing its free parameters, we test them using measurements of CMB anisotropies, SN luminosity distances, and distances inferred using the BAO feature in galaxy correlation functions. Furthermore, we assess the goodness-of-fit of CDS models by finding the maximum a posteriori (MAP), comparing the results against the common $w_0-w_a$ parameterization.

This paper is organized as follows. In Section~\ref{sec:theory} we briefly introduce the theoretical setup for the CDS scenario, discussing the modified continuity equations for dark matter and dark energy at the background and linear perturbation levels; in Section \ref{sec:eos} we define the effective and the mirage equations of state to characterize the phantom crossing in these models; in Section~\ref{sec:data} we present data analysis techniques and the methodology used in this work; in Section~\ref{sec:results} we present our results with the model constraints and best-fits obtained from the data; in Section~\ref{sec:conclusions} we summarize our findings and state our conclusions.

\section{Theory}
\label{sec:theory}

We consider dark matter ($\chi$) and dark energy ($\phi$) to be canonical scalar fields, both minimally coupled to gravity and with interactions described by a dark sector potential $V_\mathrm{DS}(\phi, \chi)$. We set the action to have the form
\begin{equation}
\begin{aligned}
S = \int d^4x \sqrt{-g}\Bigg[
    &\frac{1}{2}\planckmass R
    - \frac{1}{2}\partial_{\mu}\phi\,\partial^{\mu}\phi
    - \frac{1}{2}\partial_{\mu}\chi\,\partial^{\mu}\chi \\
    &~~~- V_\mathrm{DS}(\phi, \chi)
\Bigg],
\end{aligned}
\label{eq:action}
\end{equation}
where $g$ is the metric determinant, $M_{pl}$ is the reduced Planck mass, and $R$ is the Ricci scalar. We further assume that the potential $V_\mathrm{DS}(\phi, \chi)$ can be written as the sum of a self-interaction potential for the dark energy field $V(\phi)$ and an interaction term in which the dark matter particle mass depends on the dark energy field, $V_\mathrm{int}(\phi, \chi) = m^2(\phi) \chi^2/2$. 
Therefore, the model is characterized by two functions of the dark energy field, $V(\phi)$ and $m(\phi)$~\cite{Rosenfeld_2007}. %This class of models was studied in \cite{van_de_Bruck_2023} where conditions on the potential $V_\mathrm{int}$ were found such that the fields $\phi$ and $\chi$ are able to describe late-time dark matter and dark energy, and furthermore this model was shown to alleviate the Hubble tension in~\cite{teixeira2024alleviatingcosmologicaltensionshybrid}.

In order to describe non-relativistic CDM, we impose that the DM field $\chi$ must quickly oscillate around a minimum with respect to a Hubble time with an average equation of state $\braket{w_\chi} = 0$. Under this condition, after averaging the dark matter field over many periods of oscillation, the background equations can be written as
\begin{subequations}
\begin{align}
    \ddot \phi + 3H\dot \phi + \frac{d V}{d \phi} &= -\frac{\rho_\chi}{m}\frac{d m}{d \phi}\label{eq:motionDE}, \\
    \rho_\chi &= \rho_{\chi,i}\left(\frac{a_i}{a}\right)^3 \frac{m(\phi)}{m_i},\label{eq:motionDM}
\end{align}  
\label{eq:evolution}
\end{subequations}
where dots denote derivatives with respect to cosmic time $t$, $H = \dot{a}/a$ is the Hubble factor and $\rho_{\chi,i}$ and $m_i$ are the dark matter density and mass at an initial scale factor $a_i$. Those equations can be written as continuity equations for dark matter and dark energy,
\begin{subequations}
\begin{align}
    \dot{\rho}_\phi + 3H\rho_\phi\left(1 + w_{\phi} \right) &=-\dot{\phi}Q,\\
    \dot{\rho}_{\chi} + 3 H \rho_{\chi} &= \dot{\phi}Q,
\end{align}
\label{continuity}
\end{subequations}
where the standard equation of state is defined as $w_\phi=P_\phi/\rho_\phi = [\dot\phi^2/2 - V(\phi)]/[\dot\phi^2/2 + V(\phi)]$ and the interaction term $Q$ is written as
\begin{equation}
  Q = \rho_\chi \frac{d}{d\phi}(\ln f),
\end{equation}
where $f(\phi) = m(\phi)/m_i$.

The dark sector interactions also affect the behavior of DE and CDM perturbations. Assuming the conformal Newtonian gauge and neglecting the vector and tensor perturbations, the line element is given by
\begin{equation}
    ds^2=a^2[-(1+2\Psi)d\tau^2 + (1 - 2\Phi)\delta_{ij}dx^idx^j],
\end{equation}
where $\Psi$ and $\Phi$ are the metric perturbations. Denoting the background dark energy field as $\phi(t)$ and its perturbations as $\delta\phi(k, a)$, the perturbed Klein-Gordon and matter continuity equations are
\begin{subequations}
\begin{align}
    \begin{split}
        \delta\phi'' + 2\mathcal{H}\delta\phi &+ \left(k^2 + a^2\frac{d^2V}{d\phi^2}\right)\delta\phi = \\
        &(\Psi' + 3\Phi')\phi' - 2a^2Q\Psi + a^2\delta Q,
    \end{split} \\[1.5ex]
        \delta_\chi' = k^2v_\chi - 3\Phi' &- \frac{\phi'}{\phi}\delta_\chi
        + 2\frac{\phi'}{\phi}\Psi + a\frac{\delta Q}{\rho_\chi},
\end{align}
\end{subequations}
where the primes denote derivatives with respect to conformal time $\tau$, $\delta Q = \delta \rho_\chi \frac{d \ln f}{d \phi} + \rho_\chi\frac{d^2 \ln f}{d\phi^2}\delta\phi$ and $v_\chi$ is the bulk velocity of the dark matter fluid. The Euler equation for $v_\chi$ remains unchanged since there is no momentum transfer between the two components in this model.

\begin{comment}
\kaynan{O Rogério acha melhor colocar as equações no gauge síncrono. Eu vou deixar elas aqui, a gente decide qual forma deixar depois.}

$$ds^2 = -a^2 d\eta^2 + a^2 \left[ \left(1 - 2\epsilon \psi \right) h_{ij} + 2\epsilon \nabla_i \nabla_j E \right] dx^i dx^j$$

$$-\delta\phi'' - 2\mathcal{H} \delta\phi' - k^2 \delta\phi - 3\phi' \psi' + a^2 \delta\phi V_{,\phi\phi} + a^2 \frac{\delta Q}{\rho_\chi} = 0$$

$$\delta\rho_\chi'/\rho_\chi + 3\mathcal{H}\delta_\chi + \left( 3\psi' - k^2 E' - k^2 v_\chi \right) - a\frac{\delta Q}{\rho_\chi}  = 0$$

\kaynan{Tem algumas simplificações na equação da matéria escura que eu não fiz ainda}
\end{comment}

We investigate the dynamics using the Peebles-Ratra potential \cite{Peebles:1987ek, Ratra:1987rm}, an inverse-power law potential given by $V(\phi) = V_*(\phi/\phi_*)^{-\beta}$ that can successfully reproduce late-time acceleration. A constant potential is obtained for $\beta=0$. While a thawing behavior can happen in direct power-law potentials~\cite{Berghaus2024Quantifying}, we restrict ourselves to the inverse power-law (\textit{i.e.} $\beta > 0$), since the two classes of potentials have radically different solutions for the quintessence field evolution. We also choose a power-law parametrization for the dark matter mass with the form $m(\phi)\propto\phi^{\alpha}$.

Therefore, the parameters of this model are the field's initial value $\phi_i$ and velocity $\dot{\phi}_i$ at the initial scale factor $a_i$, the initial value of the dark matter mass $m_i$, the potential parameters $V_*$ and $\beta$ and the coupling parameter $\alpha$.
The parameters $m_i$ and $V_*$ are fixed by imposing the observed 
dark matter and dark energy abundance today.

%The parameters $\rho_{\chi,i}$ and $V_0$ are set by imposing the conditions $\rho_\chi(a=1) = \rho_\mathrm{cr}\Omega_c$ and $\rho_\phi(a=1) = \rho_\mathrm{cr}\Omega_\mathrm{DE}$, where $\rho_\mathrm{cr} = 3H_0^2/(8\pi G)$.

In the following, we will examine different combinations of free parameters, as indicated in Table~\ref{tab:model_names}. We denote the Coupled Dark Sector models with constant potential ($\beta=0$), $\alpha=1$ and zero initial field derivative as CDS, and those with the Peebles-Ratra potential as CDS-PR. We also include in the acronyms the different parameters that are being varied.
The model was implemented as a modification of the Boltzmann solver \texttt{CAMB} \cite{Lewis_2000}. The background equations~(\ref{eq:evolution}) are all implemented in conformal time $\tau$. While we present the governing equations in the Newtonian gauge for physical clarity, the numerical implementation is performed in the synchronous gauge. We use the NEWUOA~\cite{Powell2006} algorithm to find the values of $m_i$ and $V_*$ that match the current dark matter and dark energy densities.

\begin{table}
\centering
\begin{tabular}{l c c c} 
 \hline
 Name & Potential & Interaction & Parameters \\ 
 \hline
 CDS & 	$V=V_*$ & $m\propto\phi$ & $\{\phi_i\}$\\ 
 CDS-$\phi'_i$ & 	$V=V_*$ & $m\propto\phi$ & $\{\phi_i, \phi_i'\}$\\
 CDS-$\alpha$ & 	$V=V_*$ & $m\propto\phi^\alpha$ & $\{\phi_i, \alpha\}$\\
 CDS-$\alpha,\phi'_i$ & 	$V=V_*$ & $m\propto\phi^\alpha$ & $\{\phi_i, \phi_i', \alpha\}$\\
 \midrule
 PR & $V\propto \phi^{-\beta}$ & $m=m^*$ & $\{\phi_i, \beta\}$\\
 CDS-PR & $V\propto \phi^{-\beta}$ & $m\propto\phi$ & $\{\phi_i, \beta\}$\\
 CDS-PR-$\phi'_i$ & $V\propto \phi^{-\beta}$ & $m\propto\phi$ & $\{\phi_i, \phi_i', \beta\}$\\
 CDS-PR-$\alpha$ & $V\propto \phi^{-\beta}$ & $m\propto\phi^\alpha$ & $\{\phi_i, \alpha, \beta\}$\\
 CDS-PR-$\alpha,\phi'_i$ & $V\propto \phi^{-\beta}$ & $m\propto\phi^\alpha$ & $\{\phi_i, \phi_i', \beta, \alpha\}$\\
 \hline
\end{tabular}
\caption{Self-interacting potentials and free parameters used in the analysis. The term $V_*$ is the one that for which a shooting is performed in every case, while the initial dark matter density is set by $\phi_i$, we chose $\phi_*=1~M_{pl}$ as a reference and we leave $\phi'_i=0$ when it is not listed as a parameter.}
\label{tab:model_names}
\end{table}

\begin{figure}[ht]
	\centering 
	\includegraphics[width=0.48\textwidth]{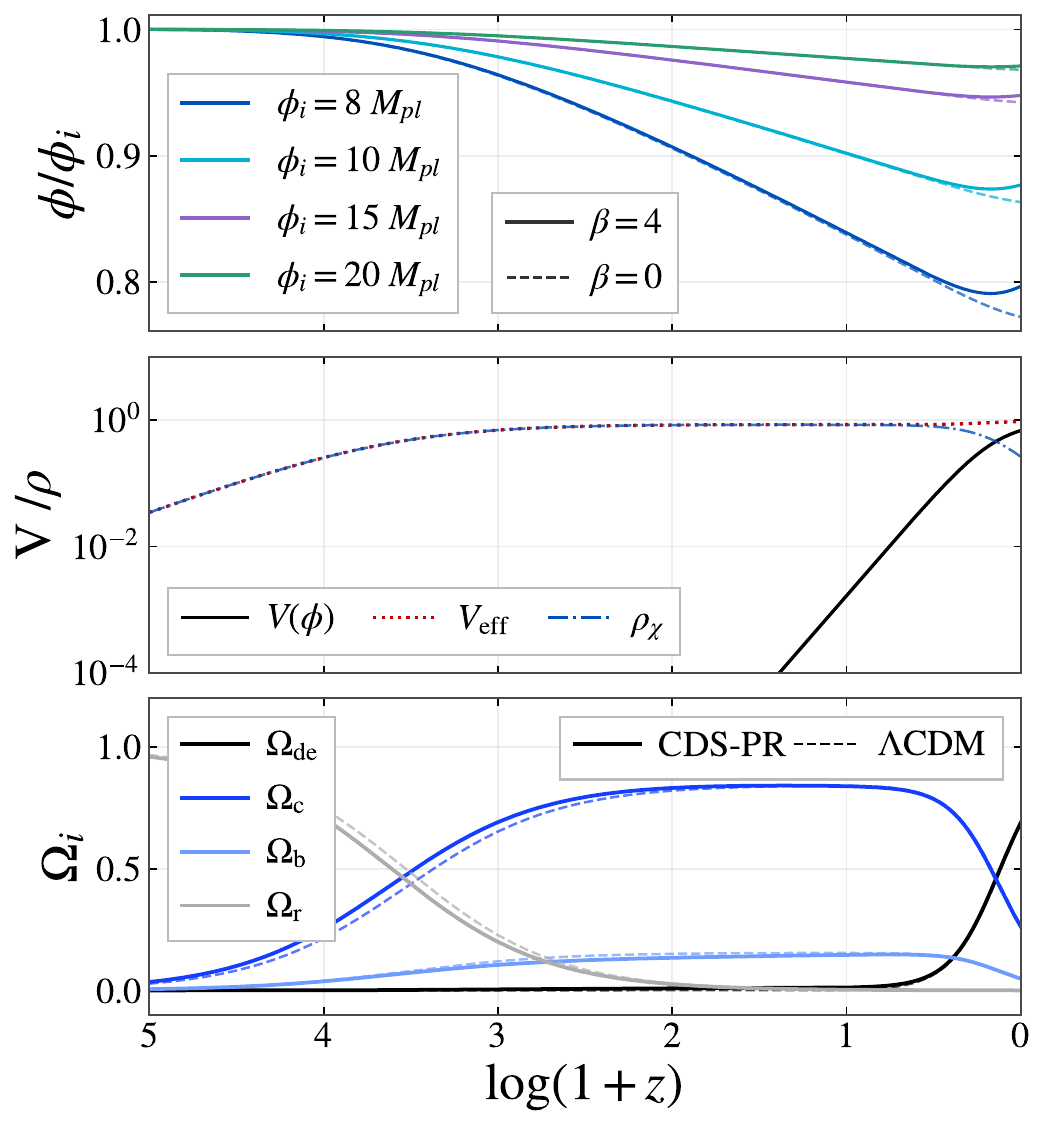}	
	\caption{Background dynamics of the model. Computations were carried out using $\alpha=1$ and $\beta=4$ for the CDS-PR case. $\phi_i=8M_{\mathrm{pl}}$ is used unless otherwise stated. \emph{Upper panel}: evolution of the normalized quintessence field as a function of redshift. The solid lines represent the CDS-PR case, while the dashed ones the CDS case. Starting the evolution with a smaller amplitude increases the field displacement both at early and late times. The $\Lambda$CDM case is recovered as $Q\to0$ and $\beta\to0$. \emph{Middle panel}: the effective potential and its components for the CDS-PR model with $\phi_i=8~M_{pl}$, normalized by the total energy density. At early times the effective potential tracks $\rho_\chi$ as the scale of $V$ is negligible, its contribution becomes ever more relevant in late-times. Lower panel: the background fractional energy density of each species. The $\Lambda$CDM case (dashed line) is compared with the CDS-PR case (solid line).}
	\label{fig:field_and_potential}%
\end{figure}

As an example, we show in the upper panel of Figure \ref{fig:field_and_potential} the solution to the background equations for several values of $\phi_i$, assuming $\dot\phi_i=0$, $\beta=4$ and $\alpha=1$. 
The smaller initial field amplitudes make the scalar field roll faster, increasing its displacement. This is due to the interaction term, which scales with $\alpha/\phi$. We also notice that the self-potential only affects the field dynamics at late times.

\begin{figure}[ht]
	\centering 
	\includegraphics[width=0.48\textwidth]{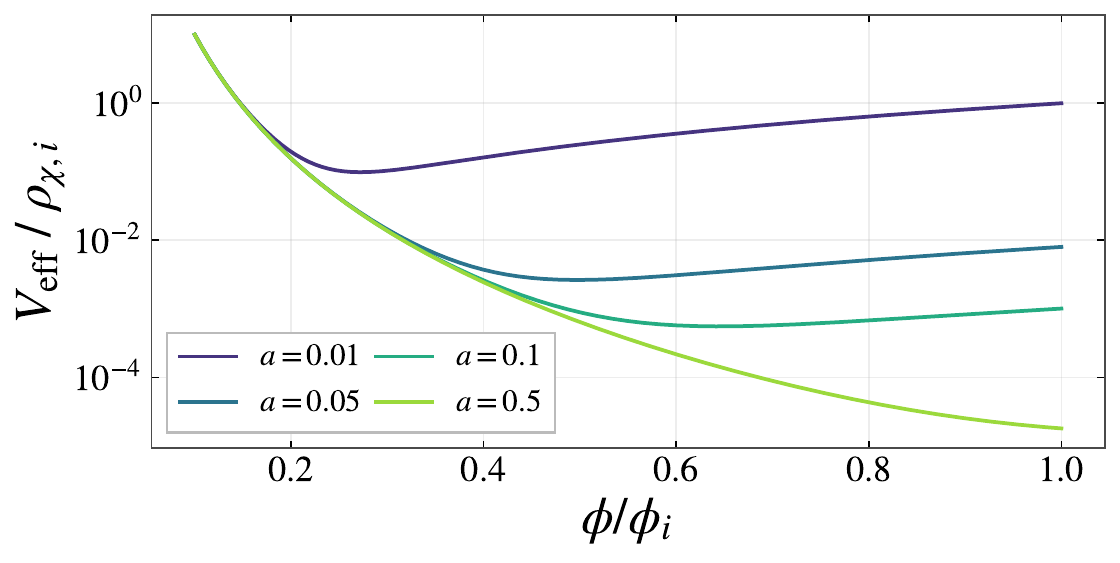}	
	\caption{Illustrative image of the effective potential for increasing values of the scale factor in the CDS-PR scenario. We've chosen $a_i=0.01$, $\alpha=2$, $\beta=6$ and $V_i=10^{-5}\rho_{\chi,i}$. }
	\label{fig:effective_potential}%
\end{figure}

The scalar field dynamics is controlled by an effective potential that can be derived from Equation~\ref{eq:motionDE} as:
\begin{equation}
    V_\mathrm{eff}=V(\phi)+\rho_{\chi}\quad .
\end{equation}
The form of the effective potential is shown in the middle panel of Figure~\ref{fig:field_and_potential} and in Figure \ref{fig:effective_potential}.
At early-times, the $\rho_\chi$ term dominates since the DE contribution is negligible.
At late-times ($z \approx 0.8$), as the dark matter density dilutes, the self-potential dominates the dynamics. This explains why the self-potential affects the field dynamics only at late times, as noticed above.
%The energy scale of $V(\phi)$ must be around $10^{-47}~\text{GeV}^4$ to reproduce the observed value of $\Omega_{DE}$ today, setting the field to a slow-roll. At late-times ($z \approx 0.8$), as the dark matter density dilutes, the bare potential dominates the dynamics, surpassing the interaction.

Although the scalar field energy density is small for redshifts $z>1$, its effect on the dark matter energy density is non-negligible even at large redshifts. This is due to modifications in the dark matter mass as the scalar field evolves. This is shown in a comparison of the normalized density of different species between CDS and $\Lambda$CDM, shown in the lower panel of Figure \ref{fig:field_and_potential}. For $\alpha>0$, the mass tends to decrease with respect to its value at recombination. Hence, in order to match the present-day dark matter density, it is necessary to have a greater CDM density in the past. This feature becomes more relevant with greater variations of the mass, and leaves an imprint on early-time observables such as the CMB, even though the dark energy density remains negligible until late-times. It is interesting to notice that this behaviour for dark matter resolves the "matter-era distance excess" that drives the tension with $\Lambda$CDM \cite{Weiner:2026sfm}.

Moreover, in the case of the Peebles-Ratra potential, it can be seen in Figure \ref{fig:effective_potential} that the effective potential has a time-dependent minimum and as the field rolls past it, it eventually stops ($\dot{\phi}=0$) and starts rolling back increasing its value. This behavior becomes more prominent as $\phi/\phi_i$ becomes smaller (see Figure \ref{fig:field_and_potential}). This type of dynamics is generically related to a phantom crossing in the mirage equation of state, as we will see in the next section.

\section{Three Dark Energy Equations of State}
\label{sec:eos}

\begin{figure}[H]
	\centering 
	\includegraphics[width=0.48\textwidth]{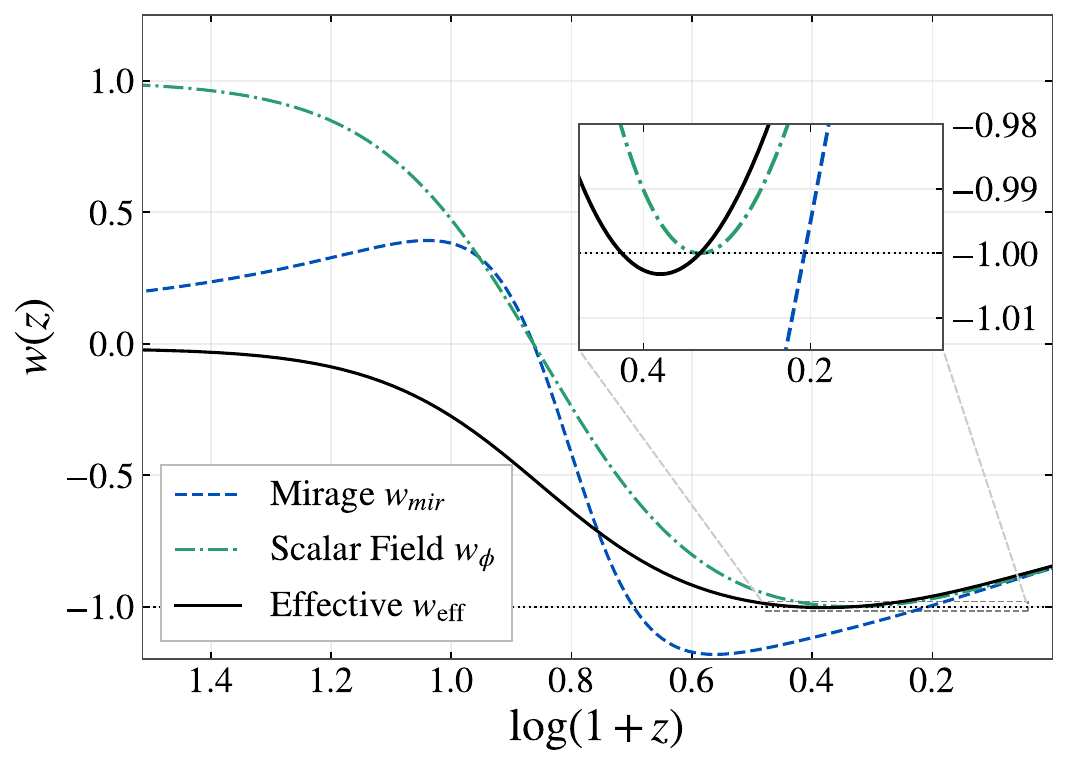}	
	\caption{The different dark energy equations of state, as a function of redshift, for $\phi_i=10~M_{pl}$, $\dot{\phi}_i=0$, $\alpha=1$ and $\beta=10$. Both the mirage and effective equation of state can phantom cross at late-times, with the canonical $w_\phi$ remains bounded $|w_\phi|\leq1$ everywhere. It is clear that the phantom crossing of the effective equation of state is much less significant than the mirage one.} 
	\label{fig:eos_1}%
\end{figure}

In order to gain more intuition about the possibility of phantom crossing in these models, one first notices that while the standard equation of state remains canonical,  we can define an interaction-dependent effective equation of state for the dark energy components $w_\mathrm{eff}$ as~\cite{Miranda2018}
\begin{equation}
    w_\mathrm{eff} = w_\phi+\frac{\dot{\phi}}{3H} \frac{Q}{\rho_\phi}=w_\phi+\frac{1}{3H} \frac{\rho_\chi}{\rho_\phi}\frac{\dot m(\phi)}{m(\phi)}
\end{equation}
in such a way that
\begin{equation}
    \dot{\rho}_\phi + 3H\rho_\phi\left(1 + w_\mathrm{eff} \right) = 0.
\end{equation}
The same parametrization can be imposed to the dark matter component, which in this case is merely a statement that its energy density does not simply scale as $a^{-3}$, but is also modulated by the dark energy field due to the mass variation. One can show analytically and numerically (see Figure \ref{fig:eos_1}) that the effective dark energy equation of state crosses into the phantom regime, which is not pathological at all: the dark energy density grows because part of it comes from the dark matter energy density. 
%One can show that an effective phantom crossing is achieved whenever the following conditions are satisfied \joao{Quem são $t_1$ e $t_2$? Não sei se esse parágrafo é muito claro} \kaynan{$t_{1,2}$ são os momentos em que acontece o phantom crossing. No modelo com $V=V_0$ dá inclusive para encontrar uma expressão analítica para eles e mostrar que $t_1\leq t_2$. Como a gente não tá muito interessado neles acho que é só tirar mesmo (até por que com $V=V(\phi)$ podem haver mais $t$'s onde $w_{eff}<-1$)}
\begin{comment}
\begin{equation}
    \dot{\phi}(t_1)=0, \quad \ddot{\phi}(t_2)=-\frac{\partial V}{\partial \phi} \label{conditions}
\end{equation}
and as long as the slope of the effective equation of state is different from zero, which is determined by the interaction term.    
\end{comment}

It is important to notice that $w_\mathrm{eff}$ should not be directly compared to
the $w_0-w_a$ best-fits resulting from the combination of CMB, BAO and SN. The reason is that in the $w_0-w_a$ case, the dark matter density scales as $a^{-3}$, which is not true for dark sector interactions, see Equation~\ref{eq:motionDM}. The fitter who is unaware of the interaction would end up fitting for the total dark sector energy,

\begin{equation}
    \rho_{\mathrm{DM}}(a)+\rho_{\mathrm{DE}}(a)=\rho_{\phi}(a)+\rho_{\chi}(a)\quad,
\end{equation}
where $\rho_\mathrm{DM}$ and $\rho_\mathrm{DE}$ are $w_0w_a$CDM parameterizations of the DE and DM densities. Therefore, this fitter finds a ``mirage" equation of state for the dark energy $w_\mathrm{mir}$ defined implicitly by~\cite{Rosenfeld_2007} 
\begin{align}\label{eq:density_mirage}
 \rho_{\phi}^0 e^{-3\int (1+w_\mathrm{eff})d(\ln{a})} +  \rho_{\chi}^0 f(\phi) a^{-3} =& \\ \nonumber
 \rho_\mathrm{DE}^0 e^{-3\int (1+w_\mathrm{mir})d(\ln{a})} +\rho_\mathrm{DM}^0 a^{-3}.
\end{align}
In other words, if interactions between dark matter and dark energy exist and are neglected in the analysis, the observer would infer a mirage equation of state that may present a phantom crossing~\cite{lapenna2026mimickingphantomdarkenergy, chen2025quintessentialdarkenergycrossing, liu2026phantommirageaxiondark}.

For illustration, we show in Figure \ref{fig:eos_1} the three different equations of state $w_{\phi}, w_\mathrm{eff}$ and $w_\mathrm{mir}$ for a CDS-PR model with initial conditions  $\phi_i=10~M_{pl}$ and $\dot{\phi}_i=0$ at $a_i = 10^{-5}$, a power-law index $\beta=10$ and interaction index $\alpha = 1$. Imposing $\rho_{\chi}^0$ = $\rho_{\mathrm{DM}}^0$ and combining eq.~(\ref{eq:density_mirage}) with the respective continuity equations for each component, one can show that the mirage equation of state can be written as \cite{chen2025quintessentialdarkenergycrossing, lapenna2026mimickingphantomdarkenergy}
\begin{align}
    w_\mathrm{mir} = \frac{w_\phi}{1+\Delta\rho_\chi/\rho_\phi} \label{apparent_w},
\end{align}
where $\Delta \rho_\chi = \rho_\chi-\rho_{\mathrm{DM}}= \rho_\mathrm{DM}(m_\chi/m_0-1)$, where $m_0$ is the present-day dark matter mass. The mirage equation of state may present a phantom crossing at late times: a simple exercise shows that, when $w_\phi<0$, $w_{\mathrm{mir}}<-1$ if $-1<\Delta \rho_\chi / \rho_\phi <-1-w_\phi$, which immediately requires that $\Delta \rho_\chi<0$ or $\phi/\phi_0<1$. From Figure \ref{fig:field_and_potential} we see that $\phi_0/\phi_i<1$ is the natural behavior for both the constant and the PB potentials. Without the addition of the potential, $\phi>\phi_0$ \emph{always}. That is, if the minimum value of $\phi$ is $\phi_{\mathrm{min}}$, then $\phi_0=\phi_{\mathrm{min}}$.

We wish to explain why the addition of a PB potential is necessary and why it must have $\beta>0$. To do so, we stress that the field dynamics control the sign of $\Delta\rho_\chi$. Without a mechanism to drive the field toward higher values (such that $\phi_0>\phi_{\mathrm{min}}$), it monotonically decreases until it reaches zero today, which causes the behaviour of $w_\mathrm{mir}$ to be drastically different from $w_0w_a$, with a very different expansion history.

%This has to be the case because

%and in that case, both models are able to cross the phantom divide regardless of the potential. However, there is an important difference: since the field dynamics control the sign of $\Delta\rho_\chi$, without a mechanism to drive the field toward higher values, once the field crosses the mirage phantom divide, it might stay there. In this case, the behavior of $w_\mathrm{mir}$ would be drastically different from $w_0w_a$, with a very different expansion history.

To clarify it, we highlight two aspects of the mirage equation of state. First, $w_\mathrm{mir} = w_\phi$ at $a = 1$, because regardless of interactions,  $\Delta\rho_\chi = 0$ today by definition. Second, \emph{if the potential is flat}, then the field asymptotically freezes, $w_\phi \to -1$, along with $w_\mathrm{eff}$ as the coupling term goes to zero at late-times as $\rho_\phi>\rho_\chi$ and $\dot \phi \to 0$. $w_\mathrm{mir}$ follows the same late-time pattern because $\Delta\rho_\chi\to0$. In that case, not only can there not be a phantom crossing, but it should also not be possible to have a similar profile for $w_{\mathrm{mir}}$ and $w_0w_a$ as $z\approx0$, because to reproduce the $w_0w_a$ curve, it is necessary that $w_{\mathrm{mir}}\to w_\phi \to w_0>-1$ as $z\to 0$.   

Upon the addition of the Peebles-Ratra potential, at late-times the effective potential becomes dominated by a steep hill, as it can be seen in Figure \ref{fig:effective_potential}. As the field climbs it, its kinetic energy decreases until its velocity changes sign and it starts rolling down, with an increase in its kinetic energy. In Figure \ref{fig:eos_1}, this regime implies that $w_\mathrm{eff}\approx w_\mathrm{mir}\approx w_\phi > -1$, which can be observed at late-times and is compatible with the $w_0w_a$ best-fits, where $w_0\approx-0.7$. The steepness of the equation of state at late-times is directly related to the steepness of the potential, $V'$, that is proportional to $-\beta/\phi$, which must be big enough to match the steepness of the $w_0w_a$ curve.

The scenario described above cannot be realized in the flat potential case. Therefore, for the field to go through a phantom phase \emph{and} reproduce the present-day densities for $\rho_\chi$ and $\rho_\phi$ while matching the background expansion of $w_0w_a$, it is \emph{necessary} that at some point its velocity changes sign, inverting the flow of energy from $\rho_\phi$ to $\rho_\chi$, such that $\Delta \rho_\chi$ grows until it becomes zero at $z=0$. This phenomenology is the main reason why the region $\beta<0$ is excluded from our analysis. 

From the discussion above, the mirage equation of state  $w_\mathrm{mir}$ 
is what should be compared to the fitted $w_0-w_a$ model.
In the following, we confront the different setups of scalar field models against the combined datasets of CMB, BAO, type Ia Supernovae and BBN, assessing the $w_\mathrm{mir}$ obtained from the CDS best-fits.

\begin{comment}
\begin{figure}[H]
	\centering 
	\includegraphics[width=0.48\textwidth]{mirage_scalar_late_time_beta10.pdf}	
	\caption{Late-time behavior of the mirage equation of state for different values of the initial field amplitude for CDS-PR1 with $\beta=10$. For greater late-time acceleration it is required that $\phi_i<10~M_{pl}$ even for such a steep potential. \kaynan{Use $w_{mir}$}} 
	\label{eos_beta10}%
\end{figure}    
\end{comment}

\section{Data Analysis}
\label{sec:data}

Our modified version of \texttt{CAMB} is integrated into the \texttt{Cocoa} pipeline~\cite{Cocoa}, which provides a distribution of \texttt{Cobaya}~\cite{Torrado_2021}. The posterior distributions were obtained using Markov Chain Monte Carlo techniques; the sampled parameters and their priors are shown in Table~\ref{priors}. When running CMB chains, we adopt a higher lower bound on the $\phi_i$ prior due to numerical instabilities at low values of $\phi_i$; this choice is sufficient to capture the best-fit model within the prior. The sum of the neutrino masses is fixed to $\sum m_\nu=0.06~\mathrm{eV}$. We use the Gelman-Rubin criterion with $|R-1| < 0.01$ to consider the chains to be converged~\cite{GelmanRubin1992}, which are then analyzed using the kernel density estimation algorithm implemented in \texttt{GetDist} \cite{Lewis_2025}. To obtain the best-fit models, we maximize the posterior probability using a simulated annealing algorithm as described in~\cite{procoli}, using 5 temperatures decreasing from 1 to $10^{-3}$. To compare the goodness-of-fit for the different models, we compute $\Delta\chi^2_\mathrm{MAP}$, the difference in log-likelihood for the maximum a posteriori parameters found from the optimization process.

\begin{table}[h] 
\centering
\begin{tabular}{l c c}
 \hline 
 Parameter & Prior\\ 
 \hline
 $\phi_i$ & $\mathcal{U}[0.1, 30]$ (geometric) | $\mathcal{U}[4, 30]$ (CMB) \\ 
 $\phi_i'$ & $\mathcal{U}[-0.3, 0.3]$\\
 $\alpha$ & $\mathcal{U}[0, 5]$\\
 $\beta$ & $\mathcal{U}[0, 10]$\\
 \hline
 $\log{(10^{10}A_s)}$ & $\mathcal{U}[1.61, 3.91]$\\
 $n_s$ & $\mathcal{U}[0.8, 1.2]$\\
 $H_0$ & $\mathcal{U}[20, 100]$\\
 $\Omega_bh^2$ & $\mathcal{U}[0.005, 0.1]$\\
 $\Omega_ch^2$ & $\mathcal{U}[0.001, 0.99]$\\
 $\tau$ & $\mathcal{U}[0.01, 0.8]$\\
 \hline
 $w_0$ & $\mathcal{U}[-3, 1]$\\
 $w_a$ & $\mathcal{U}[-5, 5]$\\
 \hline
\end{tabular}
\caption{Priors for the model parameters sampled in this work. All the priors are taken to be uniform, denoted as $\mathcal{U}[a, b]$. Nuisance parameters are sampled from the default priors specified in each likelihood of the different data sets used in our analysis.}
\label{priors}
\end{table}

The baseline datasets used in this analysis are listed below:

\begin{itemize}
    \item \textbf{Planck (CMB)}: We utilize the Cosmic Microwave Background (CMB) data from the Planck collaboration \cite{Tristram_2024} through the \texttt{Hillipop} and \texttt{Lollipop} likelihoods. The \texttt{Hillipop} likelihood consists of a high-$\ell$ ($\ell \geq 30$) TTTEEE power spectra analysis using a physical model for foreground residuals based on the PR4 (NPIPE) maps. For the low-multipole regime, we include the \texttt{Lollipop} likelihood, which focuses on the EE polarization spectra to provide a clean estimation of the reionization optical depth $\tau$. This is supplemented by the Planck PR4 lensing potential measurements, which utilize the NPIPE processing to provide an improved signal-to-noise ratio for the lensing power spectrum $C_l^{\phi\phi}$ over previous releases \cite{Carron_2022}. When analyzing geometric datasets only, we include a gaussian prior on the angular scale of the sound horizon at recombination,
    \begin{equation}
        100\theta_* = 1.04110 \pm 0.00053.
    \end{equation}

    \item \textbf{DESI DR2 (BAO)}: We include the latest Baryon Acoustic Oscillation (BAO) measurements from the Dark Energy Spectroscopic Instrument Data Release 2 \cite{Abdul_Karim_2025}. This dataset utilizes multiple tracers across a wide redshift range, including the Bright Galaxy Sample (BGS), Luminous Red Galaxies (LRGs), Emission Line Galaxies (ELGs), and Quasars (QSOs), extending up to the Lyman-$\alpha$ forest ($z \sim 4.1$). The likelihood accounts for the expansion history by measuring the transverse comoving distance $D_M(z)/r_d$, the Hubble distance $D_H(z)/r_d$, and, where applicable, the spherically averaged distance $D_V(z)/r_d$, all scaled by the sound horizon at the drag epoch $r_d$ \cite{Adame_2025}. The full covariance matrices are implemented to account for the correlations between these distance indicators across different redshift bins.

    \item \textbf{Pantheon+ (SN)}: For the late-time expansion history, we employ the Pantheon+ compilation of Type Ia Supernovae (SNIa). This catalog consists of over 1,700 light curves from 1,550 distinct SNeIa, covering a redshift range of $0.001 < z < 2.26$ \cite{Brout_2022, Scolnic_2022}. Since we combine this dataset with CMB, we do not use the absolute supernova magnitude calibration from the SH0ES collaboration \cite{Riess_2022}.

    \item \textbf{Big Bang Nucleosynthesis (BBN)}: when analyzing geometric data only, we impose a gaussian prior on the physical baryon density obtained from the \texttt{PRyMordial} code \cite{Burns:2023sgx}:
    \begin{equation}
        \Omega_b h^2 = 0.02218 \pm 0.00055.
    \end{equation}
\end{itemize}

We analyze two combinations of the datasets described above: BAO+SN+BBN+$\theta_*$ and CMB+BAO+SN.

\section{Results and Discussion}
\label{sec:results}

\begin{figure*}
    \centering
    \includegraphics[width=0.8\linewidth]{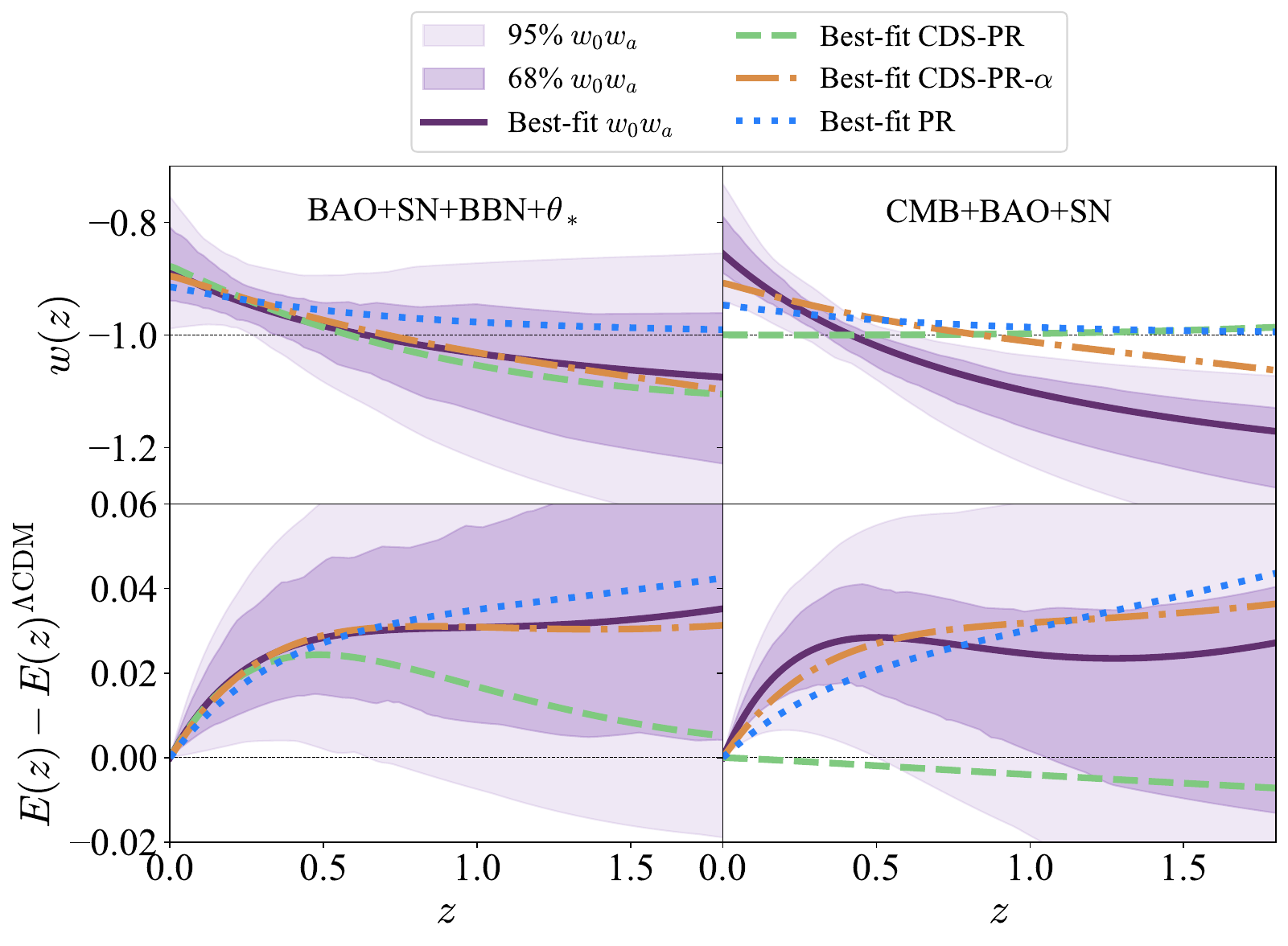}
    \caption{Constraints on the dark energy equation of state (top panels) and difference in normalized Hubble parameter $E(z)=H(z)/H_0$ relative to the best-fit $\Lambda$CDM model (bottom panels). The left panels shows constraints using the BAO+SN+BBN+$\theta_*$ dataset combination, while the right panels shows results for the CMB+BAO+SN combination. In both cases we display the $w_0w_a$CDM constraints as well as the best fits for the PR, CDS-PR, and CDS-PR-$\alpha$ models.}
    \label{fig:w_mir}
\end{figure*}

\begin{comment}
\begin{figure*}
    \centering

    \begin{subfigure}[t]{0.49\linewidth}
        \centering
        \includegraphics[width=\linewidth]{Figures/constraints_expansion_geo.pdf}
        \caption{BAO+SN+BBN+$\theta_*$.}
        \label{fig:w_mir_geo}
    \end{subfigure}
    \hfill
    \begin{subfigure}[t]{0.49\linewidth}
        \centering
        \includegraphics[width=\linewidth]{Figures/constraints_expansion_cmb.pdf}
        \caption{CMB+BAO+SN.}
        \label{fig:w_mir_cmb}
    \end{subfigure}

    \caption{Constraints on the dark energy equation of state (top panel) and normalized Hubble parameter $E(z)=H(z)/H_0$ (bottom panel) assuming the $w_0w_a$ model. Panel (a) shows the BAO+SN+BBN+$\theta_*$ dataset combination, while panel (b) shows the CMB+BAO+SN combination. In both cases we also display the best fits for the PR, CDS-PR, and CDS-PR+$\alpha$ models. \joao{Juntar essas duas figuras, usar os mesmos eixos, colocar só uma legenda}}
    \label{fig:w_mir}
\end{figure*}
\end{comment}

\begin{figure}[h]
    \centering
    \includegraphics[width=1\linewidth]{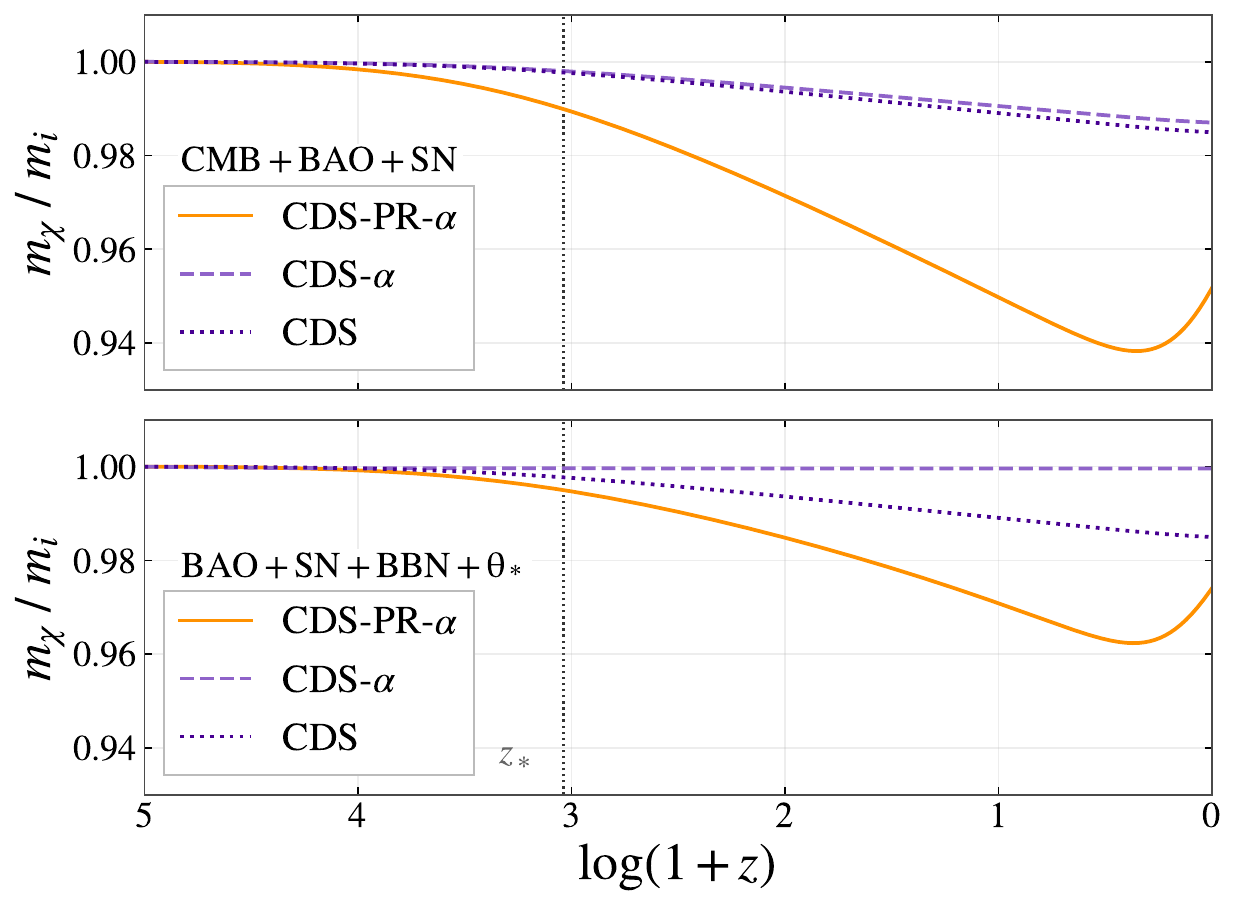}
    \caption{Best-fit evolution of the normalized dark matter mass in different models for the full dataset combination (upper panel) and for the geometric dataset alone (lower panel).}
    \label{fig:DM_mass_best_fits}
\end{figure}

\begin{figure}
    \centering
    \includegraphics[width=\columnwidth]{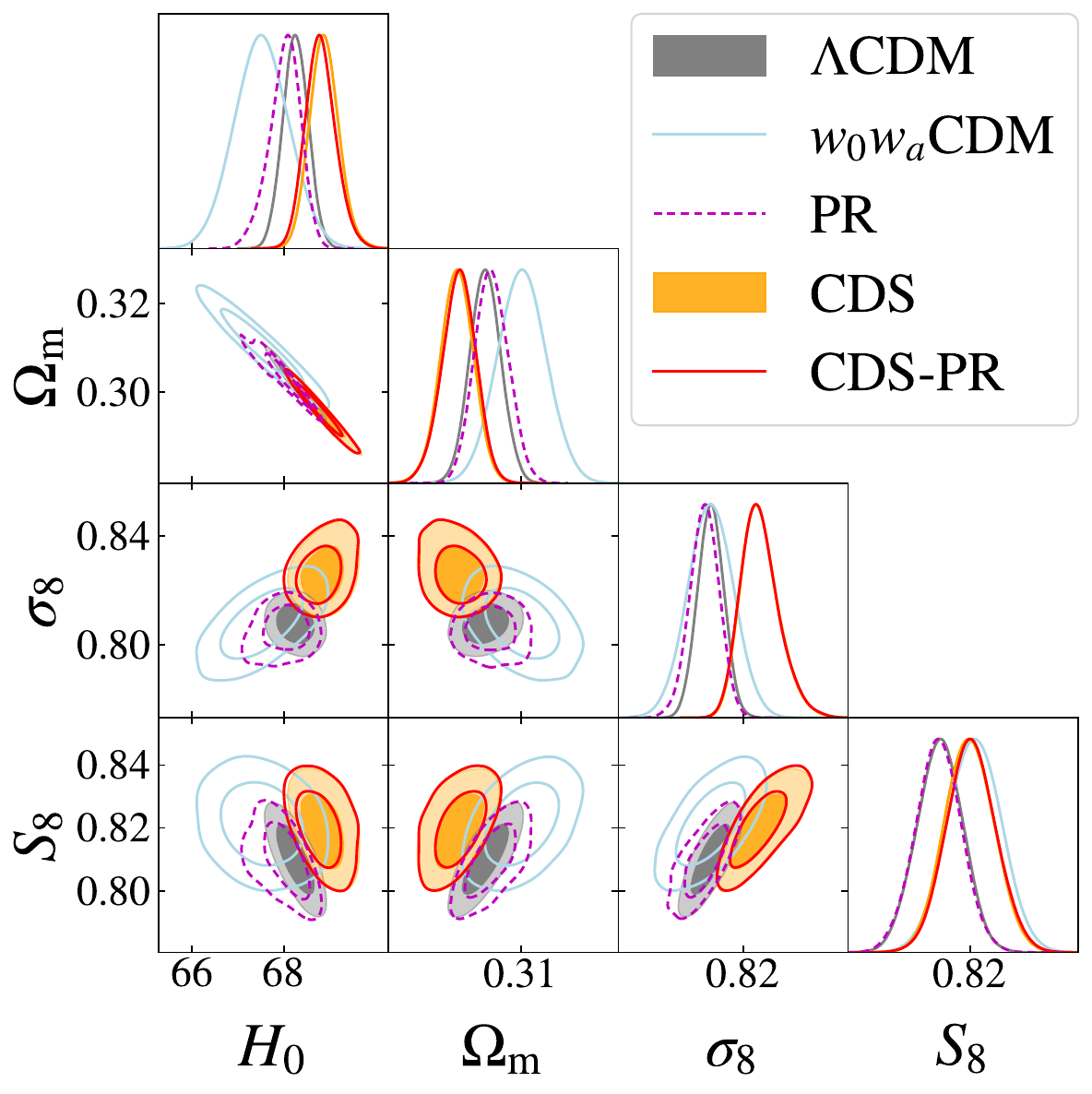}
    \caption{Posterior confidence contours (68\% and 95\%) on cosmological parameters $H_0$, $\Omega_m$, $\sigma_8$ and $S_8$ for $\Lambda$CDM, $w_0w_a$CDM and coupled quintessence models, using the dataset combination of CMB+BAO+SN.}
    \label{fig:main_dists}
\end{figure}

\begin{figure*}
    \centering
    \includegraphics[width=0.85\linewidth]{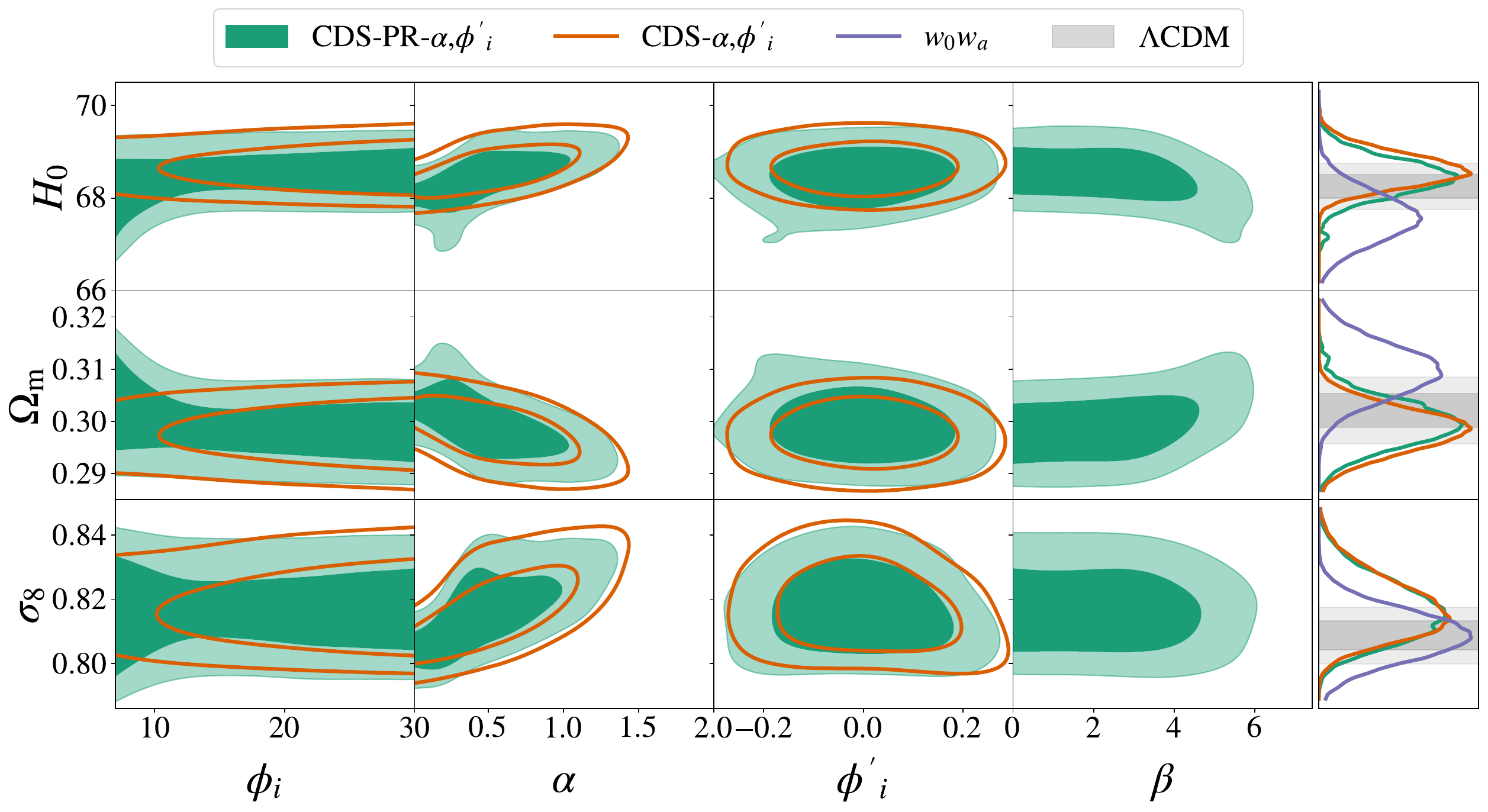}
    \caption{Marginalized 2D constraints on standard cosmological parameters as well as coupled quintessence parameters, using the dataset combination of CMB+BAO+SN. The posteriors are obtained by considering both $\alpha$ and $\phi'_i$ as free parameters.}
    \label{fig:correlations}
\end{figure*}

\begin{table*}[]
    \centering
    \small
    \renewcommand{\arraystretch}{1.3}
    \begin{tabular}{ccccccccc}
        \toprule
        Model & $\Omega_m$ & $\Omega_b$ & $H_0$ & $S_8$ & $\phi_i$ & $\alpha$ & $\beta$ & $\Delta\chi^2_\mathrm{MAP}$ \\\hline
        \multicolumn{9}{c}{BAO + SN + BBN + $\theta_*$} \\\hline
        $\Lambda$CDM & $0.299\pm 0.004$ & $0.0473\pm 0.0007$ & $68.31\pm 0.44$ &  --- & --- & --- & --- & --- \\\hline
        $w_0w_a$CDM & $0.309\pm 0.006$ & $0.0490\pm 0.0010$ & $67.25\pm 0.67$ & --- & --- & --- & --- & $-5.4$ \\\hline
        CDS & $0.295\pm 0.005$ & $0.0463^{+0.0010}_{-0.0008}$ & $68.95^{+0.48}_{-0.66}$ & --- & $> 19.5$ & --- & --- & $+0.4$ \\\hline
        CDS-$\alpha$ & $0.296\pm0.005$ & $0.0466^{+0.0010}_{-0.0007}$ & $68.78^{+0.49}_{-0.78}$ & --- & $> 17.9$ & $< 0.9$ & --- & $0$ \\\hline
        PR & $0.303^{+0.005}_{-0.006}$ & $0.0481 \pm-0.0010$ & $67.72^{+0.76}_{-0.62}$ & --- & $10.3^{+1.9}_{-8.3}$ & --- & $3.6^{+2.2}_{-1.9}$ & $-5.0$ \\\hline
        CDS-PR & $0.296\pm 0.005$ & $0.0458^{+0.0020}_{-0.0009}$ & $69.32^{+0.55}_{-1.0}$ & --- & --- & --- & $< 4.8$ & $-4.2$ \\\hline
        CDS-PR-$\alpha$ & $0.301^{+0.006}_{-0.008}$ & $0.0470 \pm 0.0010$ & $68.54^{+0.72}_{-0.93}$ & --- & --- & $< 0.6$ & $3.9^{+2.4}_{-2.1}$ & $-5.5$ \\\hline
        \multicolumn{9}{c}{CMB + BAO + SN} \\\hline
        $\Lambda$CDM & $0.302^{+0.003}_{-0.004}$ & $0.0480 \pm 0.0003$ & $68.20^{+0.29}_{-0.24}$ & $0.811\pm 0.008$ & --- & --- & --- & --- \\\hline
        $w_0w_a$CDM & $0.311\pm 0.006$ & $0.0489\pm 0.0009$ & $67.51\pm 0.61$ & $0.822\pm 0.008$ & --- & --- & --- & $-7.6$ \\\hline
        CDS & $0.296\pm 0.004$ & $0.0469 \pm 0.0004$ & $68.85\pm 0.32$ & $0.819\pm 0.008$ & $> 24.8$ & --- & --- & $-1.8$ \\\hline
        CDS-$\alpha$ & $0.298\pm 0.005$ & $0.0473^{+0.0007}_{-0.0005}$ & $68.62^{+0.36}_{-0.43}$ & $0.816\pm 0.009$ & $> 18.0$ & $0.6^{+0.3}_{-0.4}$ & --- & $-1.8$ \\\hline
        PR & $0.308 \pm 0.006$ & $0.0492^{+0.0008}_{-0.0009}$ & $67.44 \pm 0.57$ & $0.808^{+0.007}_{-0.008}$ & $9.0^{+0.9}_{-4.3}$ & --- & $4.3 \pm 0.9$ & $-2.2$ \\\hline
        CDS-PR & $0.296\pm 0.004$ & $0.0470 \pm 0.0004$ & $68.79\pm 0.33$ & $0.820\pm 0.008$ & $> 24.2$ & --- & $2.3\pm 1.1$ & $-1.7$ \\\hline
        CDS-PR-$\alpha$ & $0.301^{+0.005}_{-0.007}$ & $0.0477^{+0.0007}_{-0.0009}$ & $68.26^{+0.67}_{-0.45}$ & $0.820 \pm 0.009$ & --- & $0.49^{+0.14}_{-0.41}$ & $2.9 \pm 1.6$ & $-7.8$ \\\bottomrule
    \end{tabular}
    \caption{Marginalized constraints (mean and 68\% confidence intervals) on cosmological parameters obtained from different dataset combinations. The last column also shows the best-fit $\Delta\chi^2_\mathrm{MAP}$ for each model. Empty entries mean that the parameter is not part of this model or is unconstrained.}
    \label{tab:constraints}
\end{table*}

% The results of our MCMC analysis are summarized in the following figures and tables, providing a comparative look at the HDS models under constant and power-law (PR) potential scenarios.

Our results are summarized in Figure~\ref{fig:w_mir}, which shows the constraints on the dark energy equation of state and normalized Hubble parameter $E(z) = H(z)/H_0$ assuming the $w_0-w_a$ model and the best-fits for the PR, CDS-PR and CDS-PR-$\alpha$ models, and Table~\ref{tab:constraints}, which shows the marginalized 1D constraints and minimization results. Full contour plots as well as best-fit parameters and $\chi^2_\mathrm{MAP}$ numbers per dataset are shown in~\ref{app:chi2_per_exp}.

We start by discussing results obtained from the dataset combination of BAO+SN+BBN+$\theta_*$, in which most information comes from geometry. We first remark that, without the thawing dark energy potential, the CDS model, \textit{i.e.} without the inverse power-law potential and with $\alpha=1$, is unable to reproduce the improvement in the fit of geometric datasets, with a best-fit $\chi^2$ close to that of $\Lambda$CDM. Previous works~\cite{teixeira2024alleviatingcosmologicaltensionshybrid} have studied this model in the context of the Hubble tension and, in fact, compared to $\Lambda$CDM and $w_0w_a$CDM models, we find that both the CDS and CDS-PR models result in slightly higher $H_0$ values, relaxing the Hubble tension but not providing a natural solution. When $\alpha$ is promoted to a free parameter, we see that $\alpha = 1$ is slightly disfavored by geometric datasets.

With the thawing behavior induced by the inclusion of the potential, the quintessence models provide a similar fit to the geometric datasets than $w_0w_a$~\cite{Berghaus2024Quantifying}, with or without interactions. The inclusion of interactions with free $\alpha$ brings a slight improvement on the goodness-of-fit, reducing the $\chi^2$ by $-0.5$ compared to the case without interactions. Furthermore, Figure~\ref{fig:w_mir} shows that the CDS-PR-$\alpha$ is the model that can best reproduce the $w_0w_a$CDM expansion history, with best-fit dark energy parameters of $\phi_i \approx 2 \mathrm{M}_\mathrm{pl}$, $\alpha \approx 0.1$ and $\beta \approx 1.3$, which can be seen as the posterior peak in Figure~\ref{fig:full_triangle_geo_cds_pr}. Figure~\ref{fig:full_triangle_geo_cds_pr} also displays a degeneracy between $\phi_i$ and $\beta$, discussed previously in Section~\ref{sec:eos}: the steepness of the potential is given by $V'(\phi)/V(\phi) = -\beta/\phi$. This was further observed during the minimization procedure, as the minimum values of $\phi_i$ and $\beta$ would often vary along the degeneracy direction with small differences in $\chi^2$. We did not observe any significant effect of the $\phi'_i$ parameter in these results, even when combining with CMB data.

It has been suggested in previous works that mild incompatibilities in $\Omega_m$ between BAO and SN are alleviated by a phantom-crossing $w_0w_a$ cosmology~\cite{w0wa_omegam}. One interesting difference between the $w_0w_a$ and the CDS-PR-$\alpha$ model is that, while the former increases $\Omega_m$ by roughly $2\sigma$ with respect to $\Lambda$CDM, the latter has compatible $\Omega_m$ constraints with $\Lambda$CDM. This is also the case with the constraints with full CMB data, as we discuss below. This discrepancy, related to the "matter-era distance excess"  \cite{Weiner:2026sfm},
can be explained by dark matter interactions with other species, as the dark matter mass evolves in time. In Figure~\ref{fig:DM_mass_best_fits}, we show the evolution of dark matter mass as a function of redshift for our best-fit models. The CDS-PR-$\alpha$ model presents the largest variation in dark matter mass, which is decreased by 4-6\% from its initial value at low redshifts depending on the dataset under consideration. The inverse power-law character of the potential (see Figure~\ref{fig:effective_potential}) makes the scalar field value increase at very low redshifts, leading to a small increase of the dark matter particle mass in this period. In the absence of the PR potential, the dark matter mass evolves in essentially the same way for the CDS and CDS-$\alpha$ cases, most notably for the CMB + BAO + SN dataset combination. Furthermore, it is suggested in \cite{colgain2026predictiondesifullshapeincreasing} that the phantom crossing observed by DESI is a consequence of a trend in $\Omega_m$ across different redshift bins, where it decreases towards the past before increasing again. This behavior is qualitatively consistent with the mass variation reported in this work, as this is the profile of the dark matter mass for the best-fits with phantom crossing, as can be seen in Figures \ref{fig:w_mir} and \ref{fig:DM_mass_best_fits}.

We now turn our attention to the combination of CMB, BAO and SN datasets. While CMB is mostly insensitive to late-time smooth dark energy~\cite{planck2015, planck2018}, interactions between DE and DM can significantly impact the CMB anisotropies~\cite{van_de_Bruck_2023}. Therefore, although the CDS-PR-$\alpha$ has been successful in describing the dynamical dark energy signal from geometric-only datasets, it is important to confirm whether these results persist when including full CMB data.

The constraints on the standard cosmological parameters $H_0$, $S_8$, $\sigma_8$ and $\Omega_m$, are presented in Fig.~\ref{fig:main_dists}, where we compare the baseline $\Lambda$CDM model against the $w_0w_a$CDM, PR, CDS and CDS-PR models (see Table~\ref{tab:model_names}). Marginalized 1D constraints are shown in Table~\ref{tab:constraints}. We first notice the agreement between the CDS and CDS-PR cases, indicating that for $\alpha = 1$, adding the potential power-law index $\beta$ does not affect the cosmological constraints. Once again we notice a difference in background parameter constraints between smooth dark energy models (\textit{i.e.} the $w_0w_a$CDM and PR models) and the coupled quintessence models. While the $w_0w_a$ model shifts $H_0$ by $-1.0\sigma$ and $\Omega_m$ by $+1.3\sigma$ with respect to $\Lambda$CDM, the coupled quintessence models shift $H_0$ by $+1.6\sigma$ and $\Omega_m$ by $-1.2\sigma$. The uncoupled quintessence model produces shifts in the same directions as $w_0w_a$, although with much less significance.

As also shown in Figure~\ref{fig:main_dists}, while $w_0w_a$ does not affect $\sigma_8$ in comparison to $\Lambda$CDM, the CDS models increase $\sigma_8$ by approximately $1.9\sigma$ due to their direct effects in the dark matter background density and perturbations. Interestingly, $w_0w_a$ and CDS models have similar impacts in $S_8$, with respective increases in $S_8$ of approximately $1.4\sigma$ and $1.0\sigma$ compared to $\Lambda$CDM. These small shifts in $S_8$ persist when $\alpha$ is a free parameter. Nevertheless, the minimization results, shown in Table~\ref{tab:constraints}, imply that only the CDS-PR-$\alpha$ model can bring a noticeable improvement in $\chi^2$ comparable to the $w_0w_a$CDM model.

Figure~\ref{fig:correlations} shows the 2D marginalized constraints on cosmological parameters for the CDS-$\alpha,\phi'_i$ and CDS-PR-$\alpha,\phi'_i$ models. As noted before, when the $\alpha$ parameter is allowed to vary, the datasets prefer values below $\alpha = 1$. As expected, in the limit $\alpha \rightarrow 0$, the constraints on standard cosmological parameters $H_0$, $S_8$ and $\Omega_m$ tend to the PR constraints, which are close to the $\Lambda$CDM constraints shown as gray bands on the rightmost column.

The right panels of Figure~\ref{fig:w_mir} show the $w_0w_a$ constraints on $w(z)$ and $E(z)$ and the best-fit coupled quintessence models. The conclusions are similar to those of the geometric-only case: the coupled model with free $\alpha$ provides a quality of fit comparable to one in the $w_0w_a$ model. The best-fit parameters are similar to the case with geometric datasets only: $\phi_i = 6.9 M_\mathrm{pl}$, $\beta = 5.5$, and $\alpha = 0.3$, shown as peaks in the posterior distributions depicted in Figure~\ref{fig:full_triangle_cmb_cds_pr}. While the case without interactions only improves the fit by $\Delta \chi^2 = -2.2$ compared to $\Lambda$CDM, the CDS-PR with free $\alpha$ improves the fit by $\Delta\chi^2 = -7.8$, a similar result compared to $w_0w_a$CDM, with $\Delta\chi^2 = -7.6$. Coupled quintessence decrease the BAO and SN $\chi^2$ by $-0.3$ and $-0.4$ compared to $w_0w_a$CDM, but worsen the $\chi^2$ of the low multipole TT and lensing power spectra by $+0.6$ and $+0.2$. Having similar $\chi^2$ values, employing model comparison techniques such as the Akaike~\cite{akaike_information} or Bayesian~\cite{bayesian_information} information criteria, the $w_0w_a$CDM model is favored over CDS-PR-$\alpha$. Nevertheless, our results show that models of coupled dark sector can explain the observed deviations from $\Lambda$CDM from combinations of different datasets as a fundamental model for dark energy and dark matter.

Our results are consistent with the potential reconstructed in \cite{lapenna2026mimickingphantomdarkenergy}, where it is shown that the reported best-fit $w_0w_a$CDM expansion history can be reproduced exactly by a $\phi^{-4}$ potential, although with a different functional form of the mass function. We can also compare our results to those of  \cite{Chakraborty_2025} and \cite{gomezvalent2026constraintscoupleddarkenergy}, where their analysis favors smaller values for the power-law parameter $\beta$ in $V(\phi)$. However, they assume an exponential functional form for the interaction, and as it grows at late-times, the potential does not have to be very steep. 

If the evidence for dynamical and phantom dark energy persists or grows in the future, our results support a viable alternative to $\Lambda$CDM, motivating the search of more fundamental models where the late-time field acceleration can be derived.

\section{Summary and conclusions}
\label{sec:conclusions}

The small but intriguing deviation from a $\Lambda$CDM background expansion has been proven robust across multiple dataset combinations of CMB, BAO, type Ia supernovae, and galaxy 2-point correlation functions. The data can be satisfactorily fitted by a simple $w_0-w_a$ extension of the $\Lambda$CDM, and the best-fit values of the $w_0$ and $w_a$ parameters lead to a phantom crossing of the dark energy equation of state in the recent past.

This raises difficulties in the search for a dynamical dark energy model. Canonically normalized, minimally coupled and non-interacting scalar field models can not present phantom crossing. Some more complex models, such as k-essence models, can introduce instabilities. 

In this context, interactions between dark energy and dark matter provide a stable way to reproduce the observed phantom behavior. The core idea is that, when interactions occur, the dark matter density is also altered, and the unaware fitter would include the non-standard dark matter density evolution in the calculation of the dark energy equation of state. In this way, an apparent phantom crossing can occur even if dark energy and dark matter are not phantom.

The form of the energy-momentum exchange carries fundamental information about the nature of dark energy and dark matter fields. For instance,
a simple interaction term of the form $\phi^2\chi^2$ in the dark sector action, where $\phi$ and $\chi$ are the dark energy and dark matter fields, was studied in the context of the Hubble tension \cite{van_de_Bruck_2023,teixeira2024alleviatingcosmologicaltensionshybrid}.
In this work we extend this model in two ways: while the original work simply assumes a constant dark energy self-interaction potential, we include an inverse power-law potential of the form $V(\phi) \propto \phi^{-\beta}$, and we generalize the interaction term from $m_\chi \propto \phi$ to  $m_\chi \propto \phi^\alpha$, where $\alpha$ can smoothly transition between the original model of $\alpha = 1$ and the uncoupled quintessence case with $\alpha = 0$.

We find that the original model, while alleviating the Hubble tension, cannot explain the dynamical dark energy signal obtained in the combination of BAO and SN datasets anchored by BBN and $\theta_*$ priors, even when $\alpha$ is free. When including the dark energy potential, the quintessence model is able to provide a similar fit than $w_0w_a$ regardless of $\alpha$. The best fit occurs when $\alpha = 0.1$, with a mirage equation of state very close to that of the best-fit $w_0w_a$CDM model. When full CMB data is included, the coupled quintessence model can still provide similar results as the phenomenological $w_0w_a$ model, while the case without interactions cannot reproduce the full improvement of fit brought by $w_0w_a$CDM over $\Lambda$CDM.

Furthermore, the effect of interacting dark sector models on standard cosmological parameters such as $\Omega_m$ and $\sigma_8$ is different than the smooth dark energy $w_0w_a$ model. While CMB and galaxy correlation functions are not very sensitive to late-time dynamical dark energy, they are very sensitive to dark matter interactions that radically alter the structure formation. In order to extract full information about coupled quintessence models, a prescription for nonlinearities in the matter power spectrum is needed, requiring expensive N-body simulations to calibrate predictions. This effort can have an immense reward as interaction models remain as promising models to explain current dataset tensions, recent hints for deviations from $\Lambda$CDM, and the nature of dark matter and dark energy.

\section*{Acknowledgments}
We thank Elsa Teixeira and Carsten van de Bruck for help during the initial steps of this work, and Vivian Miranda for useful discussions. This material is based upon High Performance Computing (HPC) resources supported by the University of Arizona TRIF, UITS, and Research, Innovation, and Impact (RII) and maintained by the UArizona Research Technologies department. The work of KROP is supported by the São Paulo Research Foundation (FAPESP) under grant 2025/01714-4 and RR is partially supported by a CNPq research grant 311627/2021-8. RR thanks the Laboratório Interinstitucional de e-Astronomia (LIneA) for its support. This research was supported by resources supplied by the Center for Scientific Computing (NCC/GridUNESP) of the São Paulo State University (UNESP). 
\appendix
\setcounter{figure}{0}
\setcounter{table}{0}
\renewcommand{\thefigure}{A.\arabic{figure}}
\renewcommand{\thetable}{A.\arabic{table}}

\section{Full Contour Plots and Minimization Results}
\label{app:chi2_per_exp}
\begin{figure*}
    \centering
    \includegraphics[width=\linewidth]{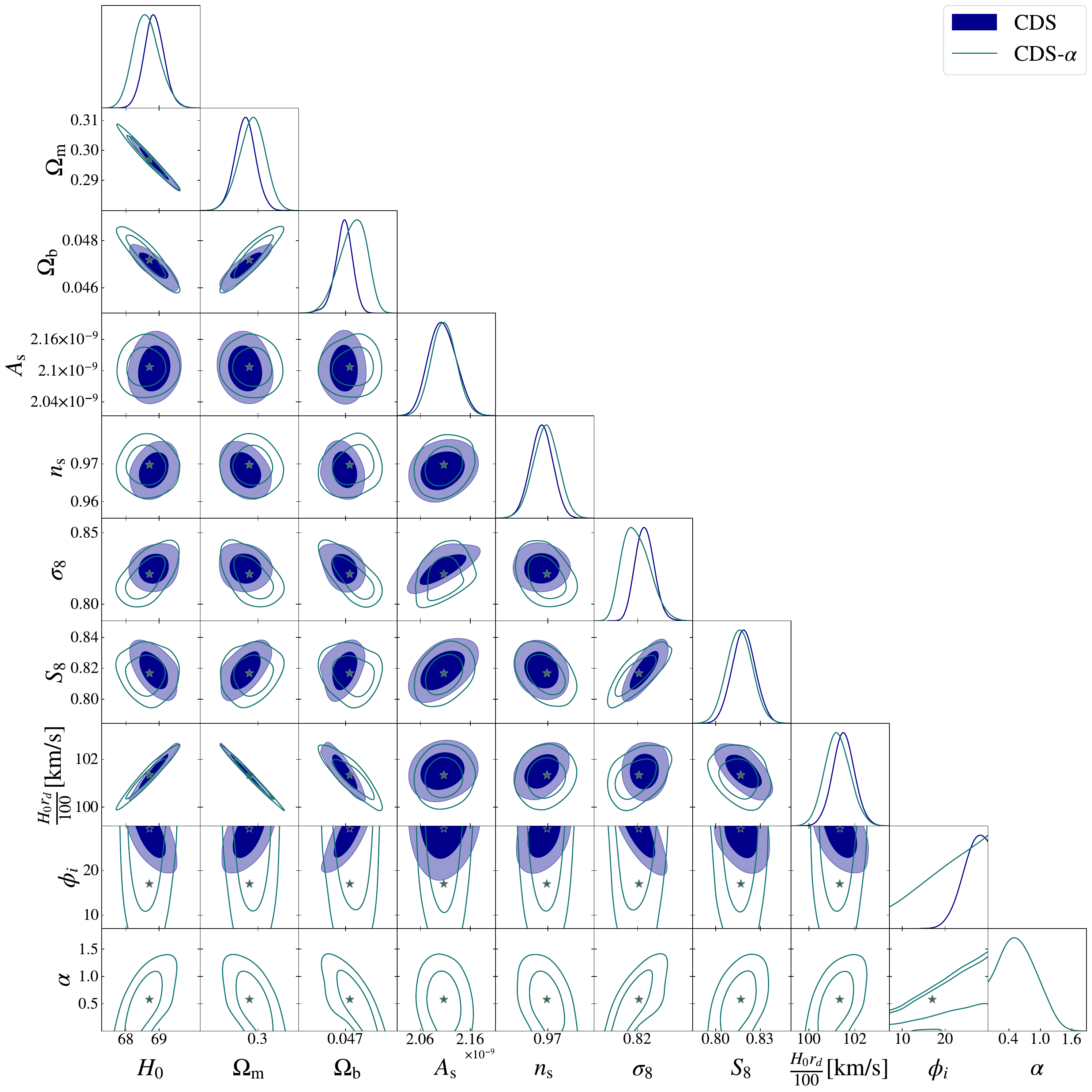}
    \caption{Marginalized 2D posterior contours for chains using the CMB+BAO+SN dataset assuming the CDS and CDS-$\alpha$ models. Colored stars denote the best-fits for each model.}
    \label{fig:full_triangle_cmb_cds}
\end{figure*}

\begin{figure*}
    \centering
    \includegraphics[width=\linewidth]{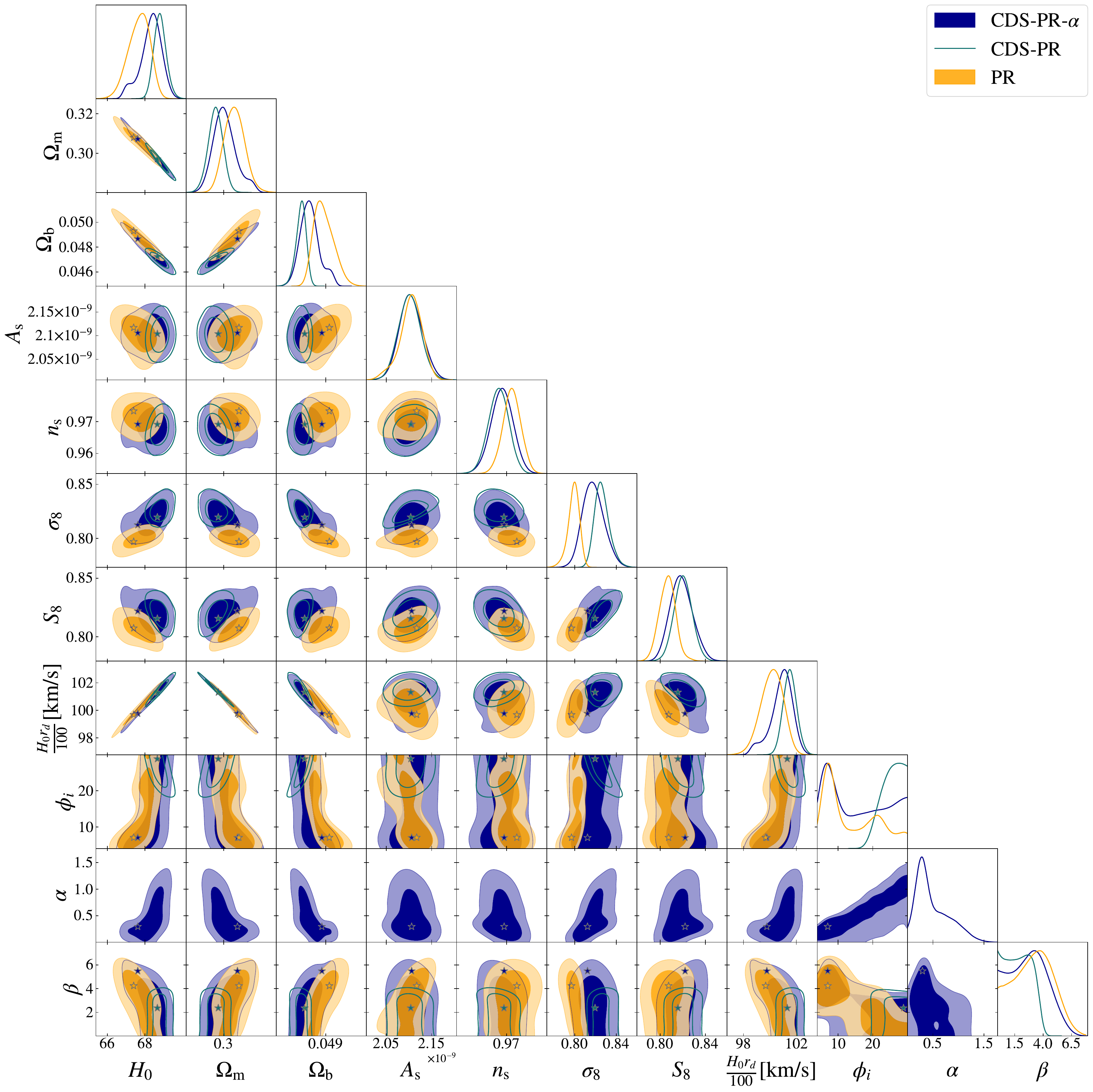}
    \caption{Marginalized 2D posterior contours for chains using the CMB+BAO+SN dataset assuming the PR, CDS-PR and CDS-PR-$\alpha$ models. Colored stars denote the best-fits for each model.}
    \label{fig:full_triangle_cmb_cds_pr}
\end{figure*}

\begin{figure*}
    \centering
    \includegraphics[width=\linewidth]{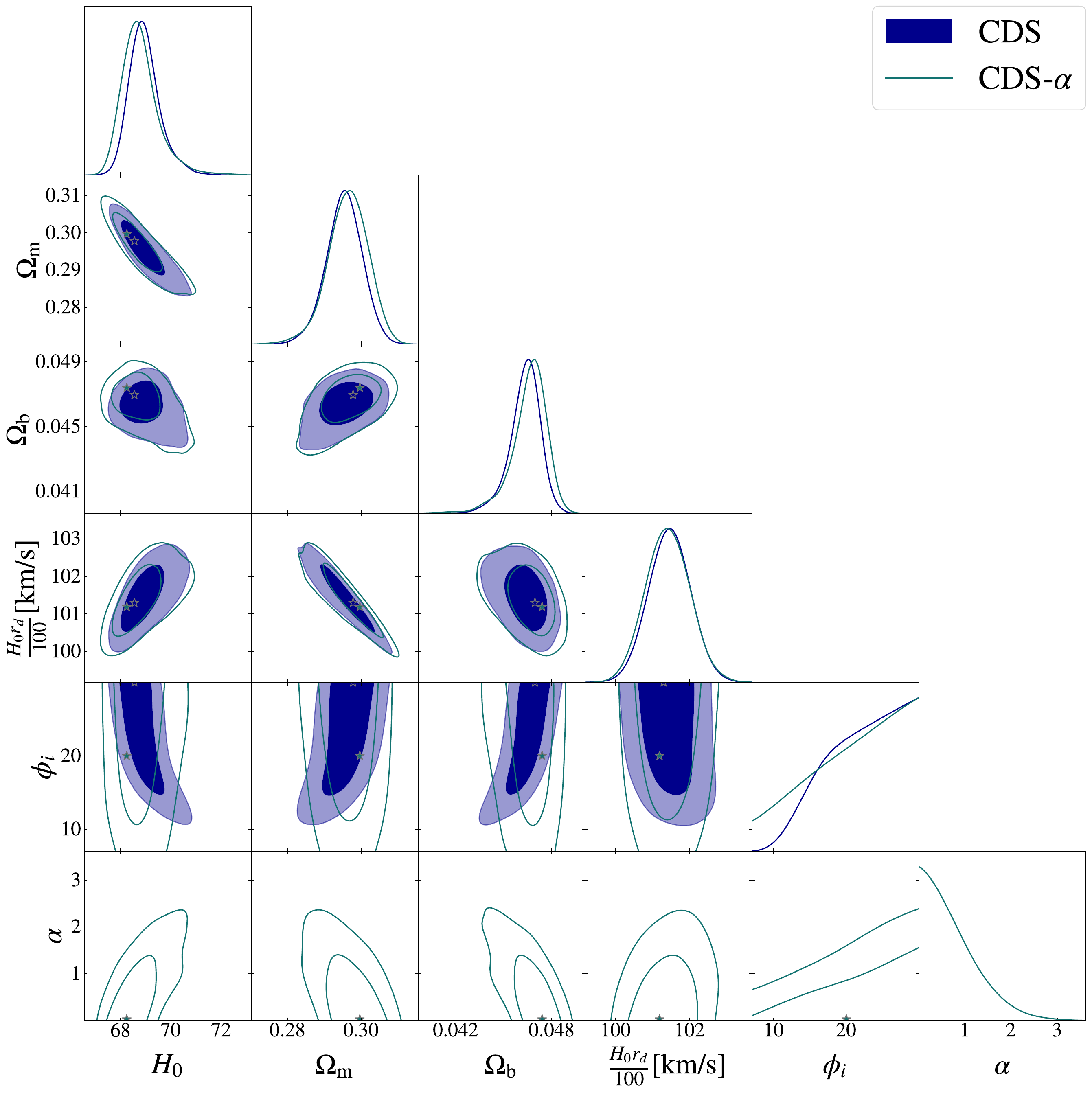}
    \caption{Marginalized 2D posterior contours for chains using the BAO+SN+BBN+$\theta_*$ dataset assuming the CDS and CDS-$\alpha$ models. Colored stars denote the best-fits for each model.}
    \label{fig:full_triangle_geo_cds}
\end{figure*}

\begin{figure*}
    \centering
    \includegraphics[width=\linewidth]{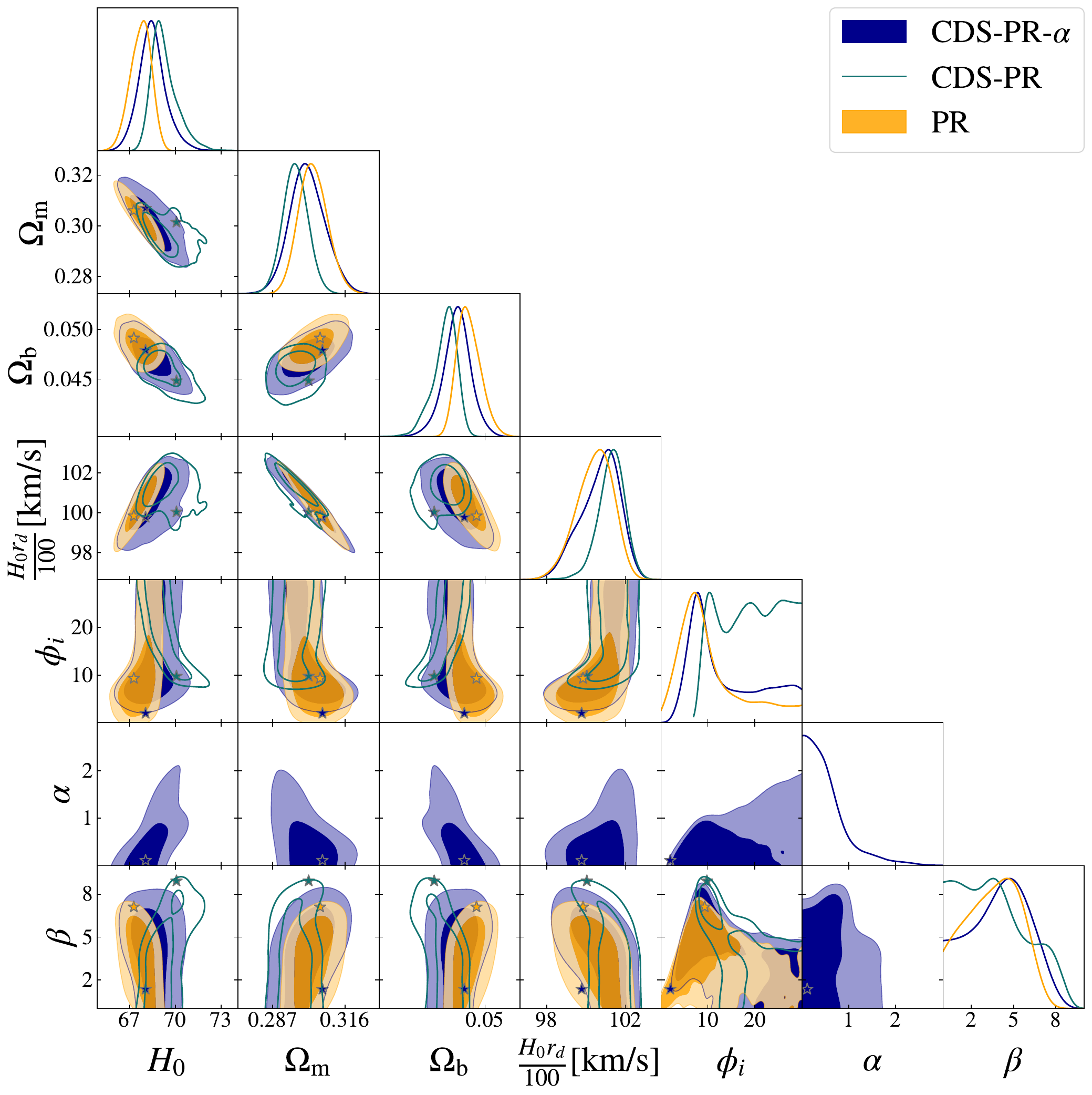}
    \caption{Marginalized 2D posterior contours for chains using the BAO+SN+BBN+$\theta_*$ dataset assuming the PR, CDS-PR and CDS-PR-$\alpha$ models. Colored stars denote the best-fits for each model.}
    \label{fig:full_triangle_geo_cds_pr}
\end{figure*}

In this Appendix, we show the full posterior contour plots from our chains in Figures~\ref{fig:full_triangle_cmb_cds}, \ref{fig:full_triangle_cmb_cds_pr}, \ref{fig:full_triangle_geo_cds} and~\ref{fig:full_triangle_geo_cds_pr}, best-fit parameter values in Tables~\ref{tab:bestfits_cmb} and \ref{tab:bestfits_geo}, and detailed results of our minimization runs separating each experiment in Table~\ref{tab:chi2-per-dataset}.


\begin{table*}
\centering
\caption{Best-fit parameters for CMB+BAO+SN chains.}
\label{tab:bestfits_cmb}
\begin{tabular}{lrrrrrrrrrrr}
\toprule
 & $\Omega_ch^2$ & $\Omega_bh^2$ & $H_0$ & $\ln(10^{10}A_s)$ & $n_s$ & $\tau$ & $w_0$ & $w_a$ & $\phi_i$ & $\alpha$ & $\beta$ \\
\midrule
$\Lambda$CDM & 0.1176 & 0.0224 & 68.25 & 3.051 & 0.972 & 0.062 & --- & --- & --- & --- & --- \\
$w_0w_a$CDM & 0.1183 & 0.0223 & 67.50 & 3.045 & 0.970 & 0.060 & -0.86 & -0.49 & --- & --- & --- \\
CDS & 0.1173 & 0.0223 & 68.71 & 3.048 & 0.970 & 0.060 & --- & --- & 29.3 & --- & --- \\
CDS-$\alpha$ & 0.1173 & 0.0223 & 68.72 & 3.048 & 0.970 & 0.060 & --- & --- & 25.1 & 0.8 & --- \\
CDS-PR & 0.1171 & 0.0223 & 68.65 & 3.046 & 0.970 & 0.060 & --- & --- & 28.9 & --- & 2.4 \\
CDS-PR-$\alpha$ & 0.1176 & 0.0223 & 67.57 & 3.047 & 0.970 & 0.060 & --- & --- & 6.9 & 0.3 & 5.5 \\
PR & 0.1169 & 0.0224 & 67.40 & 3.053 & 0.973 & 0.063 & --- & --- & 7.0 & --- & 4.2 \\
\bottomrule
\end{tabular}
\end{table*}

\begin{table*}
\centering
\caption{Best-fit parameters for BAO+SN+BBN+$\theta_*$ chains.}
\label{tab:bestfits_geo}
\begin{tabular}{lrrrrrrr}
\toprule
 & $\Omega_ch^2$ & $\Omega_bh^2$ & $w_0$ & $w_a$ & $\phi_i$ & $\alpha$ & $\beta$ \\
\midrule
$\Lambda$CDM & 0.1168 & 0.0221 & --- & --- & --- & --- & --- \\
$w_0w_a$CDM & 0.1169 & 0.0222 & -0.89 & -0.29 & --- & --- & --- \\
CDS & 0.1172 & 0.0221 & --- & --- & 30.0 & --- & --- \\
CDS-$\alpha$ & 0.1181 & 0.0221 & --- & --- & 26.8 & 0.008 & --- \\
CDS-PR & 0.1254 & 0.0220 & --- & --- & 9.8 & --- & 8.9 \\
CDS-PR-$\alpha$ & 0.1193 & 0.0222 & --- & --- & 2.0 & 0.1 & 1.3 \\
PR & 0.1156 & 0.0222 & --- & --- & 9.4 & --- & 7.1 \\
\bottomrule
\end{tabular}
\end{table*}

\begin{table*}
\centering
\caption{Chi-squared components and prior contributions for $\Lambda$CDM, $w_0w_a$CDM and CDS models.}
\label{tab:chi2-per-dataset}
\begin{tabular}{lcccccccc}
\toprule
 & $\chi^2_\mathrm{BAO}$ & $\chi^2_\mathrm{SN}$ & $\chi^2_\mathrm{CMB-high-\ell}$ & $\chi^2_\mathrm{CMB-low-\ell-TT}$ & $\chi^2_\mathrm{CMB-low-\ell-EE}$ & $\chi^2_\mathrm{CMB-lensing}$ & $\chi^2_\mathrm{BBN}$ & $\chi^2_{\theta_*}$ \\
\hline
\multicolumn{9}{c}{BAO + SN + BBN + $\theta_*$} \\\hline
$\Lambda$CDM & 10.5 & 1406.2 & --- & --- & --- & --- & 0.034 & 0.006 \\
$w_0w_a$CDM & 8.7 & 1402.6 & --- & --- & --- & --- & 0.001 & 0.000 \\
CDS & 10.6 & 1406.5 & --- & --- & --- & --- & 0.045 & 0.007 \\
CDS-$\alpha$ & 10.5 & 1406.2 & --- & --- & --- & --- & 0.037 & 0.003 \\
CDS-PR & 9.9 & 1402.7 & --- & --- & --- & --- & 0.093 & 0.023 \\
CDS-PR-$\alpha$ & 8.6 & 1402.6 & --- & --- & --- & --- & 0.000 & 0.000 \\
\hline
\multicolumn{9}{c}{CMB+BAO+SN} \\\hline
$\Lambda$CDM & 12.1 & 1405.7 & 30502.7 & 22.0 & 33.4 & 8.8 & --- & --- \\
$w_0w_a$CDM & 9.2 & 1402.9 & 30501.5 & 22.1 & 32.9 & 8.5 & --- & --- \\
CDS & 10.7 & 1406.6 & 30501.3 & 22.5 & 32.9 & 8.7 & --- & --- \\
CDS-$\alpha$ & 10.8 & 1406.6 & 30501.4 & 22.5 & 32.9 & 8.7 & --- & --- \\
CDS-PR & 10.7 & 1406.7 & 30501.4 & 22.6 & 32.8 & 8.7 & --- & --- \\
CDS-PR-$\alpha$ & 8.8 & 1402.5 & 30501.3 & 22.6 & 32.9 & 8.7 & --- & --- \\
\bottomrule
\end{tabular}
\end{table*}

\bibliographystyle{elsarticle-num}
\bibliography{references}

\end{document}